\documentclass[12pt]{article}
\usepackage{times}
\usepackage{geometry}
\usepackage[utf8]{inputenc}
\usepackage{enumitem,amssymb}
\usepackage{ragged2e}
\usepackage{graphicx}
\usepackage{fancyhdr}
\usepackage{xcolor}
\usepackage{verbatim}
\usepackage[
    natbib=true,
    style=numeric,
    sorting=none
]{biblatex}
\newlist{thematic}{itemize}{8}
\setlist[thematic]{label=$\square$}
\usepackage{pifont}

\begin{document}
\raggedright
%Quick title, feel free to suggest alternatives.  
\huge{The Unrealised Interdisciplinary Advantage of Observing High Mass Transiting Exoplanets and Brown Dwarfs}
\large \\
\vspace{4mm}
%replace \square with \boxtimes for your thematic area(s)
Aarynn L. Carter$^1$, Munazza. K. Alam$^1$, Thomas Beatty$^2$, Sarah Casewell$^3$, \hspace{5mm} Katy L. Chubb$^4$, Kielan Hoch$^1$, Nikole Lewis$^5$, Joshua D. Lothringer$^1$, \hspace{9mm} Elena Manjavacas$^6$, Sarah E. Moran$^7$, Hannah R. Wakeford$^4$ \\

\vspace{4mm}
{\small$^1$ Space Telescope Science Institute, 3700 San Martin Dr, Baltimore, MD 21218, USA \\
$^2$ Department of Astronomy, University of Wisconsin-Madison, Madison, WI 53706, USA \\
$^3$ Centre for Exoplanet Research, School of Physics and Astronomy, University of Leicester, University Road, Leicester, LE1 7RH, UK \\
$^4$ HH Wills Physics Laboratory, Tyndall Avenue, University of Bristol, BS8 1TL \\
$^5$ Department of Astronomy and Carl Sagan Institute, Cornell University, 122 Sciences Drive, Ithaca, NY 14853, USA \\
$^6$ ESA Office for Space Telescope Science Institute, San Martin Drive, Baltimore, USA \\
$^7$ Lunar and Planetary Laboratory, University of Arizona, Tucson, AZ 85721, USA \\}
\vspace{4mm}
\noindent \textbf{Thematic Areas (Check all that apply):} \linebreak $\boxtimes$ (Theme A) Key science themes that should be prioritized for future JWST and HST observations 
\linebreak $\square$ (Theme B) Advice on optimal timing for substantive follow-up observations and mechanisms for enabling exoplanet science with HST and/or JWST \linebreak
$\square$ (Theme C) The appropriate scale of resources likely required to support exoplanet science with HST and/or JWST 
 \linebreak
$\square$ (Theme D) A specific concept for a large-scale ($\sim$500 hours) Director’s Discretionary exoplanet program to start implementation by JWST Cycle 3.
%  $\square$ (Theme D) A specific concept for a large-scale ($\sim$500 hours) Director’s Discretionary exoplanet program to start implementation by JWST Cycle 3.
% \linebreak
\vspace{4mm}
\justify{
\textbf{Summary:} We advocate for further prioritisation of atmospheric characterisation observations of high mass transiting exoplanets and brown dwarfs. This population acts as a unique comparative sample to the directly imaged exoplanet and brown dwarf populations, of which a range of JWST characterisation observations are planned. In contrast, only two observations of transiting exoplanets in this mass regime were performed in Cycle 1, and none are planned for Cycle 2. Such observations will: improve our understanding of how irradiation influences high gravity atmospheres, provide insights towards planetary formation and evolution across this mass regime, and exploit JWST's unique potential to characterise exoplanets across the known population.}

\pagebreak
\justify{

\textbf{Anticipated Science Objectives:}
Comparative studies between transiting exoplanets and the more widely separated directly-imaged exoplanet and brown dwarf populations have been historically limited due to the observational biases of their different characterisation methods. Even with the advent of \textit{JWST} we cannot combine these techniques for a single target. We can, however, use the separate advantages \textit{JWST} provides to transits and direct-imaging to enable a robust comparison of atmospheric physics and chemistry between these two populations. Both directly-imaged exoplanets and brown dwarfs are being characterised through photometry \textit{and} spectroscopy for masses $\gtrsim\,$3~$M_\mathrm{Jup}$ (e.g., \cite{Wilcomb21, Hinkley23}) and temperatures as cool as 250~K. Objects with lower masses have yet to be detected using direct imaging techniques. In contrast, and despite the known transit population extending well into the brown dwarf regime, just two transiting exoplanets with masses $\gtrsim\,$3~$M_\mathrm{Jup}$ were characterised as part of Cycle 1 observations \cite{Coulombe23, Sikora21}, and none will be characterised during Cycle 2 (Fig. 1). Furthermore, those two which have been characterised have only utilised a single \textit{JWST} near-infrared observing mode each. As the catalogue of high signal-to-noise, broad wavelength coverage, \textit{JWST} spectra of directly-imaged exoplanet and brown dwarf atmospheres continues to grow, we are providing little to no opportunity for comparison to the transiting planet population. Although transmission measurements are difficult for higher mass targets, emission measurements are still achievable with just one eclipse observation (Fig. 2). With the energy budget of transit atmospheres dominated by intense irradiation, they act as a powerful contrast to imaged companions, which are dominated by the remnant heat from formation. These observations will enable comparisons between these two populations and will be able to directly probe the influence of differing radiative, advective (e.g., rotation/winds), and chemical (e.g. clouds/disequilibrium chemistry) environments in sculpting exoplanet emission spectra. Furthermore, with transits, imaged exoplanets, and isolated brown dwarfs occupying vastly different orbital separation regimes, deeper characterisation and comparison efforts will enable first-order explorations of their formation (e.g., core accretion vs. gravitational instability \cite{Molliere22}) and evolutionary histories. With giant exoplanets likely exerting a significant influence on companion exoplanets in their systems, insights into their formation and evolution may also inform the broader understanding of planetary system architectures. At a more fundamental level, these observations are a critical step towards characterising transiting exoplanet atmospheres across the breadth of the known population, which, albeit ambitious, is a goal that is well with the reach of \textit{JWST} capabilities and lifetime. 

% I think the below is only important for 500 hr DDT ideas. 
%  \vspace{0.2cm}

% \textbf{Urgency}: Your text goes here.  

% \textbf{Risk/Feasibility}: Your text goes here.

% \textbf{Timeliness}: Your text goes here.

% \textbf{Cannot be accomplished in the normal GO cycle}:  Your text goes here.

\pagebreak
\begin{figure}[!htb]
   \centering
   \includegraphics[width=0.85\linewidth]{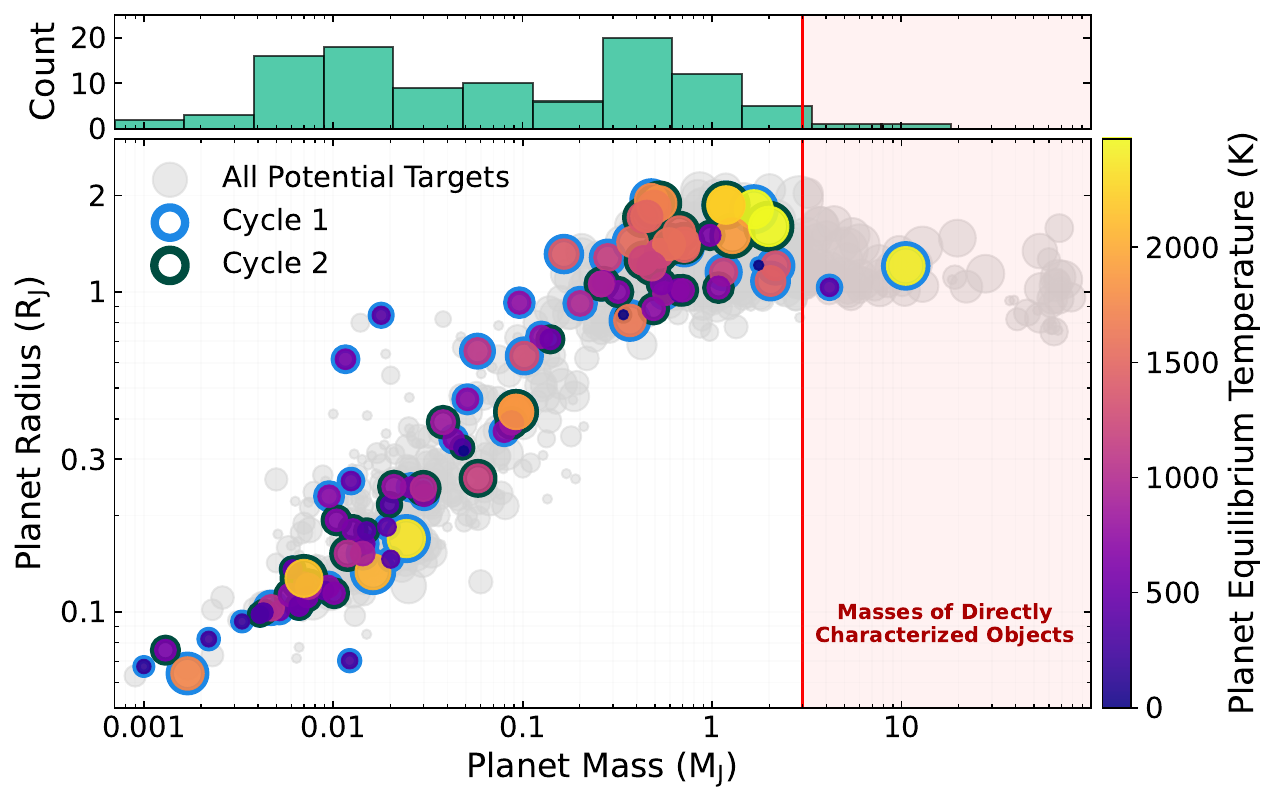} 
    \vspace{-6mm}
    \caption{The population of transiting exoplanets and brown dwarfs, as obtained from TEPCat \cite{Southworth2011}, marker size corresponds to equilibrium temperature. Targets being observed in Cycle 1 and/or 2 are highlighted accordingly, their temperature is further indicated by their color, and the histogram shows the number of targets as a function of mass. The red line indicates the lowest mass objects being characterized through direct techniques to date.} \vspace{-6mm}
\end{figure}
\vspace{-1mm}
\begin{figure}[!htb]
   \centering
   \includegraphics[width=0.95\linewidth]{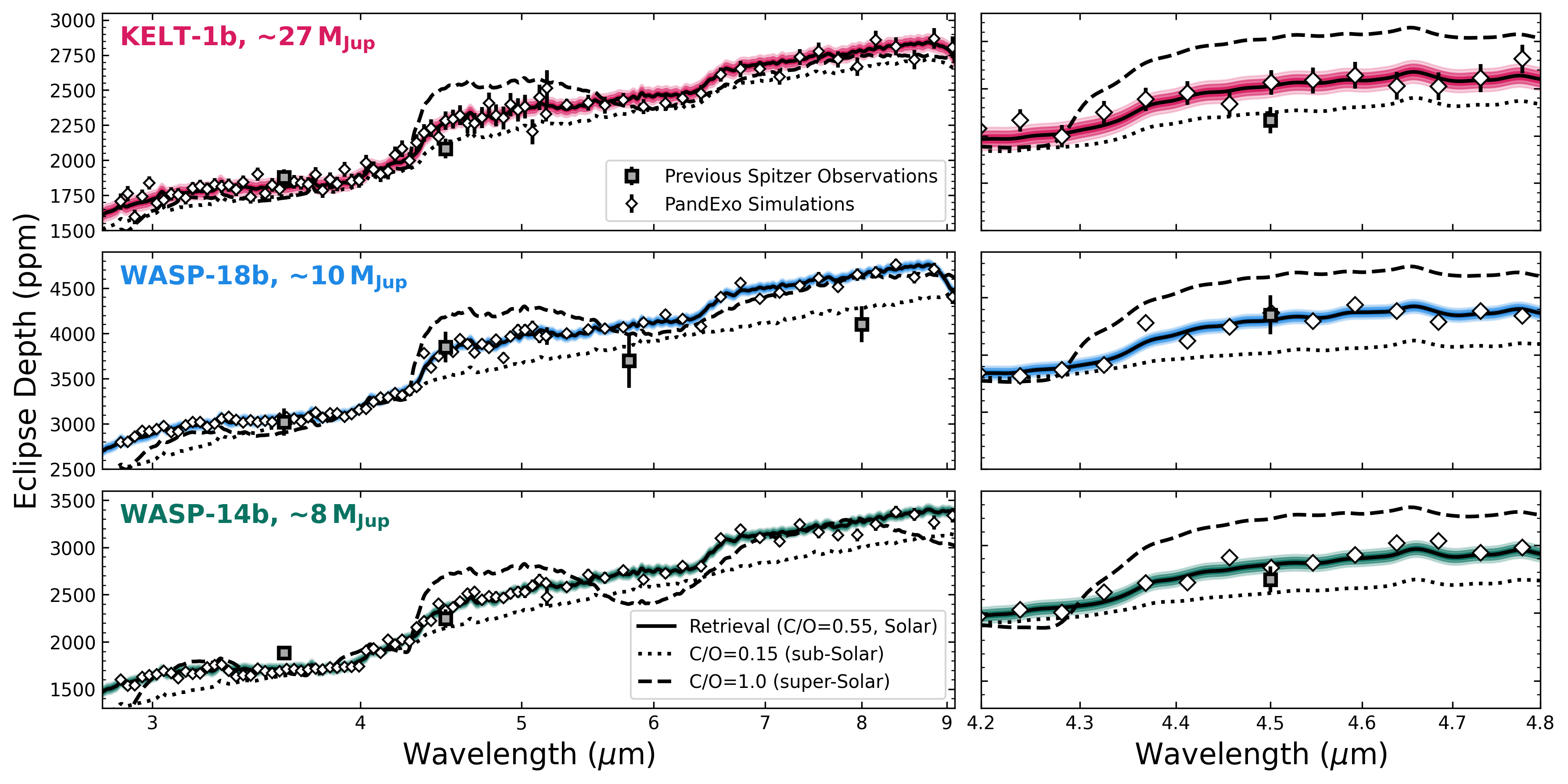} 
    \vspace{-6mm}
    \caption{Solar metallicity models of high mass targets. Simulated \texttt{PandExo} \cite{Pandexo} data (white diamonds) are shown alongside archival Spitzer data (grey squares). Data are generated from a solar C/O ratio forward model, at resolution $R$=50 for NIRSpec G395M and $R$=20 for MIRI LRS. The median retrieved model (solid line), 1, 2, and 3$\sigma$ confidence intervals (shaded regions), and sub-solar/super-solar (dotted/dashed lines) are also shown. A zoomed in portion of each simulation is displayed on the right, some errors are too small to be seen.} \vspace{-6mm}
\end{figure}
\pagebreak
}
\printbibliography
\end{document}